\documentstyle[epsfig,indentfirst,bm,12pt]{article}
\begin{document}
\title{\Large{\bf{The extraction of  nuclear sea quark distribution and energy loss effect in
Drell-Yan experiment}}}
\author{C.-G.  Duan$^{1,4}$ \footnote{\tt{ E-mail:duancg$@$mail.hebtu.edu.cn}},
 N.  Liu$^{1,2}$,  Z.-Y.  Yan$^{3,4}$}

\date{}

\maketitle \noindent {\small 1.Department of Physics, Hebei Normal
University, Shijiazhuang 050016, China}\\
{\small 2.College of Mathematics and Physics, Shijiazhuang University of Economics, Shijiazhuang 050031, China}\\
{\small 3.Department of Applied Physics, North China Electric Power
University, Baoding 071003, China}\\
{\small 4.CCAST(World Laboratory), P.O.Box8730, Beijing 100080,
China}

\baselineskip 9mm \vskip 0.5cm
\begin{abstract}

The next-to-leading order and leading order analysis are performed
on the differential cross section ratio from Drell-Yan process.  It
is found that the effect of next-to-leading order corrections can be
negligible on the differential cross section ratios as a function of
the quark momentum fraction in the beam proton and the target nuclei
for the current Fermilab  and future lower beam proton energy. The
nuclear Drell-Yan reaction is an ideal tool to study the energy loss
of the fast quark moving through cold nuclei. In the leading order
analysis, the theoretical results with quark energy loss are in good
agreement with the Fermilab E866 experimental data on the Drell-Yan
differential cross section ratios as a function of the momentum
fraction of the target parton. It is shown that the quark energy
loss effect has significant impact on the Drell-Yan differential
cross section ratios. The nuclear Drell-Yan experiment at current
Fermilab and future lower energy proton beam can not provide us with
more information on the
 nuclear sea quark distribution.



\end{abstract}

\vskip 0.5cm

\noindent {\bf 1 Introduction }

\hspace{0.5cm} In 1983, the European Muon Collaboration(EMC)
$^{[1]}$ reported that the ratios of nuclear structure functions for
the iron and deuteron nuclei is not equal to unity, but a function
of the Bjorken scaling variable $x$ in charged lepton-nucleus deep
inelastic scattering. It shows that the parton distributions are
modified in the nuclear environment because the structure function
describes the quark momentum distributions in bound nucleons. From
then on, the quark and gluon distributions in hadrons and nuclei
have been one of the most active frontiers in nuclear physics and
particle physics. The nuclear parton distribution directly affects
the interpretation of the data collected from the nuclear reactions
at high energies$^{[2]}$. The precise nuclear parton distribution
function should be very important in finding the new physics
phenomena and determining the electro-weak parameters, neutrino
masses and mixing angles in neutrino physics.

\hspace{0.5cm} In early 1970, S.Drell and T.M.Yan first studied  the
production of large-mass lepton pairs from hadron-hadron inelastic
collisions, which is so-called Drell-Yan process$^{[3]}$.  According
to the parton model, the process is induced by the annihilation of a
quark-antiquark pair into a virtual photon which subsequently decays
into a lepton pair. The nuclear Drell-Yan process of proton-nucleus
collisions therefore is closely related to the  quark distribution
functions in nuclei. It is further natural to expect that the
Drell-Yan reaction  is a complementary tool to probe the structure
of hadron and nuclei.

\hspace{0.5cm} In 1990, Fermilab Experiment772(E772) $^{[4]}$made
the measurement of the nuclear dependence of the Drell-Yan process
by using 800GeV protons bombarding on D, C, Ca, Fe, and W nuclei.
Muon pairs were recorded in the range $4GeV\leq M \leq 9GeV$ and $M
\geq 11GeV$. The covered kinematical ranges were $0.1< x<0.3$, where
$x$ is the momentum fraction of the target parton. Their aim is to
investigate the modification of the quark structure in nuclei. The
theoretical models, which can well describe the EMC effect in the
charged lepton-nuclei deep inelastic scattering, were used to fit
the observations on the Drell-Yan differential cross section ratios.
It is found that  some of the theoretical models overestimate the
nuclear Drell-Yan ratios from Fermilab E772, such as pion-excess
model$^{[5]}$ and quark-cluster model$^{[6]}$. It indicated that
whether there is another nuclear effect apart from the nuclear
effects on the parton distributions as in charged lepton-nucleus
deep inelastic scattering.

\hspace{0.5cm} In 1999, Fermilab Experiment866(E866) $^{[7]}$
performed the precise observation of the ratios of the Drell-Yan
cross section per nucleon for an 800GeV proton beam incident on Be,
Fe and W target. The Drell-Yan events extend over the ranges $4GeV<
M < 8.4GeV$, $0.01< x_2 <0.12$, $0.21< x_1 <0.95$ and $0.13< x_F
<0.93$, where $x_{1(2)}$ and $x_F$ are the momentum fraction of the
beam parton (the target nuclear parton) and Feynman scaling
variable, respectively. The extended kinematic coverage of E866
significantly increases its sensitivity to the nuclear shadowing
effect and the quark energy loss. This is the first experiment on
the energy loss of the projectile particle moving through cold
nuclei.  The E866 compared their experimental data with the results
from leading-order Drell-Yan calculations by using the EKRS nuclear
parton distributions$^{[8]}$ together with the MRST parton
distribution functions$^{[9]}$. It is shown that the energy loss
effect can be negligible.

\hspace{0.5cm} In previous works$^{[10]}$, we investigated the
Drell-Yan differential cross section ratios as the function of the
momentum fraction of the beam parton from E866 data in the framework
of Glauber model by means of EKRS and HKM nuclear parton
distribution functions$^{[11]}$. We found that the theoretical
results with energy loss of the beam proton are in good agreement
with the Fermilab E866 experiment by means of HKM nuclear parton
distributions. However, the calculated results without energy loss
can give good fits by using EKRS nuclear parton distribution
functions. Furthermore, we introduced two typical kinds of quark
energy loss parametrization, i.e.the linear and quadratic quark
energy loss with the  average path length of the incident quark in
the nucleus A. The nuclear dependence of the Drell-Yan production
cross sections were calculated by combining the quark energy loss
effect with the EKRS, HKM and HKN$^{[12]}$ nuclear parton
distribution. The $\chi^2$ global fit to the Drell-Yan differential
cross section ratios indicated that the theoretical results without
energy loss effects agree very well with the E866 experimental data
by taking advantage of the EKRS nuclear parton distribution
functions. If employing the HKM nuclear parton distribution
function, the
 results with energy loss effect are in good agreement
with the Fermilab E866 data. In addition, the results with HKM are
most near to those with HKN. It is demonstrated that at current
Fermilab incident proton energy, we can not distinguish between the
linear and quadratic dependence of the quark energy loss. Further
experiments are needed about nuclear Drell-Yan reactions with a
lower energy incident proton. Using the values of quark energy loss
from a fit to E866 experimental data, the prospects are given for
the lower energy proton beam bombarding  deuteron and tungsten. It
is shown that these future experiments can give valuable insight in
the enengy loss of fast quark propagating through cold nuclei. In
the analysis above, we performed the leading order calculations of
the Drell-Yan differential cross sections. As is well known, QCD
corrections can alter quite significantly the cross sections at a
hadronic collider. Thus, these may have serious bearing on the
discovery potential of the Drell-Yan reactions, in which the leading
order (LO) results may seriously underestimate the cross sections.
This has led to the incorporation of the next-to-leading order
(NLO)results. As for the nuclear parton distributions, EKRS, HKM and
HKN nuclear parton distributions are obtained by  Eskola et
al.$^{[8]}$, Hirai et al.$^{[11,12]}$ with the global analysis of
the relative experimental data, respectively. It is noticeable that
HKM use only the charged lepton-nucleus deep inelastic scattering
experimental data, HKN and EKRS  employ  E772 and E866 nuclear
Drell-Yan reaction data in order to pin down  the nuclear antiquark
distribution in the small x region. In this report, we will explore
the effect of NLO correction on the Drell-Yan differential cross
section ratios, and the influence  of quark energy loss on the
Drell-Yan differential cross section ratios as a function of the
momentum fraction of the target parton.

\hspace{0.5cm} This paper is organized as follows. In sect.2, a
brief formulism for the differential cross section  in the Drell-Yan
process is presented. The effect of NLO correction  is given  on the
Drell-Yan differential cross section ratios in sect.3. The influence
of quark energy loss  is given  on the Drell-Yan differential cross
section ratios as a function of the momentum fraction of the target
parton in sect.4, and the summary is given in sect.5.

\vskip 0.5cm

{\bf 2 Brief formulism for differential cross section in Drell-Yan
reaction}

\hspace{0.5cm} In the Drell-Yan process$^{[3]}$, the perturbative
QCD leading order contribution is quark-antiquark annihilation into
a lepton pair. The annihilation cross section can be obtained from
the $e^{+}e^{-}\rightarrow\mu^{+}\mu^{-}$ cross section by including
the color factor $\frac{1}{3}$ with the charge $e^{2}_{f}$ for the
quark of flavor $f$.
\begin{equation}
   \frac{d\hat{\sigma}}{dM}=\frac{8\pi\alpha^2}{9M}e^2_f\delta(\hat{s}-M^2),
\end{equation}
where $\sqrt{\hat{s}}=(x_1x_2s)^{1/2}$, is the center of mass system
(c.m.system) energy of $q\bar{q}$ collision, $x_1$(resp.$x_2$)is the
momentum fraction carried by the projectile (resp.target) parton,
$\sqrt{s}$ is the center of mass energy of the hadronic collision,
and $M$ is the invariant mass of the produced dimuon. The hadronic
Drell-Yan differential cross section is then obtained from the
convolution of the above partonic cross section with the quark
distributions in the beam and  target,
\begin{equation}
 \frac{d^2\sigma^{DY}}{dx_1dx_2}=\frac{4\pi\alpha^2}{9sx_1x_2}
 \sum_{f}e^2_f[q^p_f(x_1,Q^2)\bar{q}^A_f(x_2,Q^2)
 +\bar{q}^p_f(x_1,Q^2)q^A_f(x_2,Q^2)],
\end{equation}
where $\alpha$ is the fine-structure constant, the sum is carried
out over the light flavor $f=u,d,s$, and $q^{p(A)}_{f}(x,Q^2)$ and
${\bar q}^{p(A)}_{f}(x,Q^2)$ are the quark and anti-quark
distributions in the proton (nucleon in the nucleus A).

\hspace{0.5cm} In addition to  the leading order Drell-Yan term, the
contributions are needed from  $q\bar{q}$ annihilation processes(
$q\bar{q}\rightarrow \gamma^{*}+g$) and gluon Compton scattering(
$q+g\rightarrow \gamma^{*}+q$)$^{[13]}$ in the  perturbative QCD
next-to-leading order, which are denoted as Ann and Comp,
respectively. The contribution from the order-$\alpha_s$
annihilation graphs is
\begin{eqnarray}
 \frac{d^2\sigma^{Ann}}{dx_1dx_2}
   &=&
 \int^1_{x_1}dt_1\int^1_{x_2}dt_2[\frac{d^2\hat{\sigma}^{Ann}}{dt_1dt_2}
 \sum_{f}e^2_f q^p_f(t_1,Q^2)\bar{q}^A_f(t_2,Q^2) \nonumber\\
 & &+\frac{d^2\hat{\sigma}^{Ann}}{dt_1dt_2}(t_1\leftrightarrow t_2)\sum_{f}e^2_f \bar{q}^p_f(t_1,Q^2)q^A_f(t_2,Q^2)],
\end{eqnarray}
with
\begin{eqnarray}
 \frac{d^2\hat{\sigma}^{Ann}}{dt_1dt_2}
 &=&\sum_{f}\frac{8\alpha^2\alpha_s(Q^2)}{27Q^2}\delta(t_1-x_1)\delta(t_2-x_2)[1+\frac{5}{3}\pi^2
 -\frac{3}{2}\ln\frac{x_1x_2}{(1-x_1)(1-x_2)}\nonumber\\
 & &+2\ln\frac{x_1}{1-x_1}\ln\frac{x_2}{1-x_2}]\nonumber\\
 & &+\sum_{f}\frac{8\alpha^2\alpha_s(Q^2)}{27Q^2}\delta(t_2-x_2)[\frac{t^2_1+x^2_1}{t^2_1(t_1-x_1)_{+}}
     \ln\frac{2x_1(1-x_2)}{x_2(t_1+x_1)}\nonumber\\
 & &
 +\frac{3}{2(t_1-x_1)_{+}}-\frac{2}{t_1}-\frac{3x_1}{t^2_1}]+(1\leftrightarrow2)\nonumber\\
& &+\sum_{f}\frac{16\alpha^2\alpha_s(Q^2)}{27Q^2}
    [\frac{(\tau+t_1t_2)[\tau^2+(t_1t_2)^2]}{(t_1t_2)^2(t_1+x_1)(t_2+x_2)[(t_1-x_1)(t_2-x_2)]_{+}}\nonumber\\
& & -\frac{2\tau(\tau+t_1t_2)}{t_1t_2(t_1x_2+t_2x_1)^2}].
\end{eqnarray}
The contribution from the Compton scattering graphs is
\begin{eqnarray}
 \frac{d^2\sigma^{Comp}}{dx_1dx_2}
   &=&
 \int^1_{x_1}dt_1\int^1_{x_2}dt_2
 \{\frac{d^2\hat{\sigma}^{Comp}}{dt_1dt_2}
  \sum_{f}e^2_f g^p(t_1,Q^2)[q^A_f(t_2,Q^2)+\bar{q}^A_f(t_2,Q^2)] \nonumber\\
 & &+\frac{d^2\hat{\sigma}^{Comp}}{dt_1dt_2}(t_1\leftrightarrow t_2)
  \sum_{f}e^2_f [q^p_f(t_1,Q^2)+\bar{q}^p_f(t_1,Q^2)]g^A(t_2,Q^2)\},
\end{eqnarray}
where $g(t,Q^2)$is the gluon distribution in the beam proton and
target nuclei, and
\begin{eqnarray}
 \frac{d^2\hat{\sigma}^{Comp}}{dt_1dt_2}
 &=&\sum_{f}\frac{2\alpha^2\alpha_s(Q^2)}{9Q^2}\delta(t_2-x_2)
  [\frac{x^2_1+(t_2-x_1)^2}{2t^3_1}\ln\frac{2x_1(1-x_2)}{x_2(t_1+x_1)}\nonumber\\
  & & +\frac{1}{2t_1} -\frac{3x_1(t_1-x_1)}{t^3_1}]\nonumber\\
 & &+\sum_{f}\frac{2\alpha^2\alpha_s(Q^2)}{9Q^2}
   [\frac{x_2(\tau+t_1t_2)[\tau^2+(\tau-t_1t_2)^2]}{t^3_1t^2_2(t_1x_2+t_2x_1)(t_2+x_2)(t_2-x_2)_{+}}\nonumber\\
 & &
 +\frac{\tau(\tau+t_1t_2)[t_1t^2_2x_1+\tau(t_1x_2+2t_2x_1)]}{(t_1t_2)^2(t_1x_2+t_2x_1)^3}].
\end{eqnarray}
Therefore, to the next leading order, the differential cross section
in Drell-Yan reaction can be written  as
\begin{equation}
 \frac{d^2\sigma^{NLO}}{dx_1dx_2}=\frac{d^2\sigma^{DY}}{dx_1dx_2}+\frac{d^2\sigma^{Ann}}{dx_1dx_2}
   +\frac{d^2\sigma^{Comp}}{dx_1dx_2}.
\end{equation}
With  calculating the integral of the differential cross section
above in leading order (Eq.2) and next-to-leading order (Eq.7), the
Drell-Yan production cross section is given by
\begin{equation}
 \frac{d\sigma}{dx_{1(2)}}=\int dx_{2(1)}\frac{d^2\sigma}{dx_1dx_2}.
\end{equation}

\vskip 0.5cm

{\bf 3 The influence of QCD correction on the ratio of Drell-Yan
cross section }

\hspace{0.5cm}In the Drell-Yan reaction experiments, the ratios are
measured  of  Drell-Yan cross sections on two different nuclear
targets bombarded by proton,
\begin{equation}
R_{A_{1}/A_{2}}(x_{1(2)})=\frac{d\sigma^{p-A_{1}}}{dx_ {1(2)}}
/{\frac {d\sigma^{p-A_{2}}}{dx_{1(2)}}}.
\end{equation}
In order to explore the influence of QCD correction, we introduce
the ratios,
\begin{equation}
R^{NLO/DY}_{A_{1}/A_{2}}(x_{1(2)})=R^{NLO}_{A_{1}/A_{2}}(x_{1(2)})
/R^{DY}_{A_{1}/A_{2}}(x_{1(2)}),
\end{equation}
where$R^{NLO}_{A_{1}/A_{2}}(x_{1(2)})$ and
$R^{DY}_{A_{1}/A_{2}}(x_{1(2)})$ are the ratios of Drell-Yan
differential cross section with QCD correction and with only leading
order contribution, respectively.

\hspace{0.5cm} By  taking  advantage  of  the HKM  cubic type of
nuclear parton distribution functions $^{[11]}$, the ratios
$R^{NLO/DY}_{Fe/Be}(x_{1(2)})$ and $R^{NLO/DY}_{W/Be}(x_{1(2)})$ for
proton  incident on Be, Fe, W targets are calculated at the Fermilab
800GeV proton beam energy in the range $ 4<M<8 GeV$. It is found
that the differential cross section ratios in the next-to-leading
order are almost identical to those in the leading order. The
similar results are given for the lower energy proton bombarding
deuterium and tungsten  at the Fermilab Main Injector(FMI,120GeV
proton beam)$^{[14]}$ and the Japan Hadron Facility(JHF,50GeV proton
beam)$^{[15]}$. Therefore, it can be concluded that the QCD
correction is negligible in the nuclear Drell-Yan reactions for the
current Fermilab  and lower energy proton beam . The production of
lepton pairs in proton-nucleus collisions, the nuclear Drell-Yan
process, is one of most powerful tools to probe the propagating of
quark through cold nuclei. The experimental study of the relatively
low energy nuclear Drell-Yan process can give valuable insight in
the quark energy loss dependence on the medium size. Furthermore,
the influence of quark energy loss can be investigated on the
Drell-Yan differential cross section ratios as a function of the
momentum fraction of the target parton in the leading order.

\vskip 0.5cm

{\bf 4 The impact of quark energy loss on the Drell-Yan differential
cross section ratios as a function of the momentum fraction of the
target parton }

\hspace{0.5cm} The Fermilab E866 collaboration reported their
measurement of the differential cross section ratios
$R_{Fe/Be}(x_{2})$ and $R_{W/Be}(x_{2})$ for an 800GeV proton beam
bombarding Be, Fe and W nuclei$^{[7]}$. By combining HKM cubic type
of nuclear parton distribution$^{[11]}$  with the quark energy loss,
the global $\chi^2$ analysis to the E866 experimental data are
performed in the perturbative QCD leading order. We introduce two
typical kinds of quark energy loss expressions. One is rewritten as
\begin{equation}
\Delta x_1= {\alpha}\frac{<L>_A}{E_p},
\end{equation}
where $\alpha$ denote the  incident quark energy loss per unit
length in nuclear matter, $<L>_A$ is the average path length of the
incident quark in the nucleus A, $E_p$ is the energy of the incident
proton. The average path length is employed using the conventional
value, $<L>_A=3/4(1.2A^{1/3)}$fm. In addition to the linear quark
energy loss rate, another one is expressed   as
\begin{equation}
\Delta x_1= {\beta}\frac{<L>^2_A}{E_p},
\end{equation}
i.e., the quark energy loss is quadratic with the path length. With
considering the  quark energy loss in nuclei, the incident quark
momentum fraction can be shifted from $x'_1=x_1+\Delta x_1$ to $x_1$
at the point of fusion.

\hspace{0.5cm} If the quark energy loss do not put in, the obtained
$\chi^2$ value is $\chi^2=48.06$ for the 16 total data points. The
$\chi^2$ per degrees of freedom is given by $\chi^2/d.o.f.= 3.00$.
It is apparent that theoretical results without energy loss effects
do not significantly agree with the E866 experimental data on the
ratios $R_{A_1/A_2}(x_{2})$. After adding the fast quark energy
loss, the $\chi^2$ per degrees of freedom is  $\chi^2/d.o.f.= 1.12$
for the linear quark energy loss formula with  $\alpha=1.27$, and
the $\chi^2$ per degrees of freedom is given by $\chi^2/d.o.f.=
1.13$ for the quadratic quark energy loss expression with
$\beta=0.19$. By using the parameter values in the linear and
quadratic expressions obtained by fitting the E866 experimental data
on the ratios $R_{A_{1}/A_{2}}(x_{1})$, the $\chi^2$ per degrees of
freedom are $\chi^2/d.o.f.= 1.69$ with $\alpha=1.99$ and
$\chi^2/d.o.f.= 1.65$ with  $\beta=0.29$, respectively. The
calculated results with the linear and quadratic energy loss
expression
 are shown in Fig.1 and Fig.2. which is the Drell-Yan cross
section ratios for Fe  to Be and   W to Be as functions of $x_2$,
respectively. The solid curves are the ratios with only the nuclear
effect on the parton distribution as in deep inelastic scattering,
the dotted and dash curves respectively correspond to the linear
energy loss with $\alpha=1.27$ and $\alpha=1.99$ in Fig.1, and the
quadratic energy loss with $\beta=0.19$ and $\beta=0.29$ in Fig.2.
From comparison with the experimental data, it is shown that our
theoretical results with energy loss effect are in good agreement
with the Fermilab E866. Meanwhile, It is noticeable that the values
of the parameter $\alpha$ ( or $\beta$) are different by means of
the global $\chi^2$ analysis to the E866 experimental data on the
ratios $R_{A_{1}/A_{2}}(x_{1})$ and $R_{A_{1}/A_{2}}(x_{2})$. The
results may be  originated from the experimental precision. If the
experimental data are sufficiently precise, the values of the
parameter $\alpha$ (or $\beta$) in the quark energy loss expression
should be the same for fitting the ratios $R_{A_{1}/A_{2}}(x_{1})$
or $R_{A_{1}/A_{2}}(x_{2})$ from the nuclear Drell-Yan experiment.

\hspace{0.5cm} In order to clarify the impact of quark energy loss
on the Drell-Yan differential cross section ratios as a function of
the momentum fraction of the target parton, the ratios on
$R_{A_{1}/A_{2}}(x_{2})$ without quark energy loss to those with
linear quark energy loss  are calculated and tabulated in Table 1.
for the kinematic ranges covered by the E866 experiment. The similar
results can be obtained for the quadratic quark energy loss. It is
shown that the quark energy loss effect has  obvious influence on
the Drell-Yan differential cross section ratios
$R_{A_{1}/A_{2}}(x_{2})$. As for the ratios $R_{Fe/Be}(x_{2})$, the
variation is approximately $2\%$ to $4\%$. The extent to which the
ratios $R_{W/Be}(x_{2})$ vary is roughly $4\%$ to $8\%$. It can be
deduced that the ratios of Drell-Yan differential cross section for
nuclei A versus deuterium, $R_{A/D}(x_{2})$, are affected largely
because of the quark energy loss effect.  In the global analysis of
nuclear parton distribution functions, the Drell-Yan data are taken
mainly to determine the sea quark modification in the small $x$
region. It is obvious that, considering the existence of quark
energy loss, the application of nuclear Drell-Yan data is remarkably
subject to difficulty in the constraints of the nuclear antiquark
distribution.

\tabcolsep0.5cm
\begin{table}
\caption{the ratios of $R_{A_{1}/A_{2}}(x_{2})$ without quark energy
loss to those with linear quark energy loss.}
\begin{center}
\begin{tabular}{ccccccc}\hline
     \multicolumn{2}{c}{$x_2$}   & 0.03 & 0.05 & 0.07 & 0.09 & 0.12 \\\hline
    $\alpha(1.27)$ &$Fe/Be$      & 1.018  & 1.017 &1.019  & 1.022  & 1.027 \\
                   &$W/Be$ &      1.038 & 1.036 & 1.040& 1.045 & 1.055\\ \hline
   $\alpha(1.99)$ &$Fe/Be$      & 1.029  & 1.027 & 1.030  & 1.035  & 1.042 \\
                   &$W/Be$ &      1.061 & 1.058 & 1.064& 1.072 & 1.088\\ \hline
\end{tabular}
\end{center}
\end{table}

\hspace{0.5cm} The ratios of Drell-Yan differential cross section
for tungsten versus deuterium, $R_{W/D}(x_{2})$, are also calculated
at 50GeV and 120GeV proton beam by means of the HKM cubic type of
nuclear parton distribution functions $^{[11]}$ and two kind of
quark energy loss expressions. It is indicated that the theoretical
results with quark energy loss effect deviate  significantly those
with only the nuclear effects on the structure function. As an
example, Figure 3 shows the ratios $R_{W/D}(x_{2})$ with and without
linear quark energy loss  at 50GeV and 120GeV proton beam,
respectively. The kinematic ranges cover  $ 4<M<8 GeV$ in order to
avoid contamination from charmonium decays. In these calculation,
the energy loss per unit length $\alpha$ equal $1.27GeV/fm$ and
$1.99GeV/fm$ from a good fit to E866 data. Therefore, the Drell-Yan
experiment at a lower energy proton beam do not provide us with more
information on the nuclear sea quark distribution, especially for
large-$x$ region.

\hspace{0.5cm} In addition, we calculate the ratios $R_{W/D}(x_{2})$
with the linear and  quadratic energy loss formula in the 50GeV and
120GeV proton beam bombarding deuterium and tungsten target. It is
presented that the results with linear quark energy loss are almost
identical to those with quadratic energy loss. The ratios
$R_{W/D}(x_{2})$ do not determine whether the energy loss is linear
or quadratic with the path length. However, The ratios
$R_{W/D}(x_{1})$ can easily distinguish between $L$ and $L^2$
dependence of quark energy loss$^{[10]}$. As an example, the Fig.4
shows the ratios $R_{W/D}(x_{2})$ of the Drell-Yan cross section per
nucleon for an 120GeV proton beam incident on D and W target,  where
the solid and dotted lines correspond to a quadratic energy loss of
$\beta=0.19GeV/fm^2$ and to a linear energy loss of
$\alpha=1.27GeV/fm$, respectively.

\vskip 0.5cm

{\bf 5 Concluding remarks }

\hspace{0.5cm} As a summary, the next-to-leading order and leading
order analysis are performed on the differential cross section
ratios from nuclear Drell-Yan process. The calculated results
indicated that the QCD correction can be ignored in the nuclear
Drell-Yan reactions for the current Fermilab  and lower energy
proton beam. Based on this view, the nuclear Drell-Yan process is
one of most powerful tools to probe the propagating of quark through
cold nuclei. The experimental study in the relatively low energy
nuclear Drell-Yan process can give valuable insight in the quark
energy loss dependence on the medium size. Furthermore, we have made
a leading-order analysis of E866 data on  the Drell-Yan differential
cross section ratios as a function of the momentum fraction of the
target parton  by taking into account of the energy loss effect of
fast quarks. It is found that the theoretical results with quark
energy loss are in good agreement with the Fermilab E866 experiment.
The quark energy loss effect has  obvious impact on the Drell-Yan
differential cross section ratios $R_{A_{1}/A_{2}}(x_{2})$. The
nuclear Drell-Yan experiment at current Fermilab and future lower
energy proton beam can not provide us with more information on the
extraction of nuclear sea quark distribution. In fact, by means of
the structure function $xF_3(x,Q^2)$ in neutrino deep inelastic
scattering only, the nuclear modifications to the valence quark
distribution can very precisely be determined in the medium and
large $x$ regions. With the structure functions $F_2(x,Q^2)$ from
the neutrino and charged-leptons deep inelastic scattering off
nuclei, the nuclear modifications to the sea quark distribution in
the medium and large $x$ regions would be pinned down$^{[16]}$. The
nuclear sea quark distribution in the small-$x$ regions can be
relatively determined from the constraints of momentum conservation
and the structure function in small-$x$ range. We desire to operate
precise measurements of the experimental study in the relatively low
energy nuclear Drell-Yan process. These new experimental data can
shed light on the energy loss of fast quark propagating in a cold
nuclei.

{\bf Acknowledgement:} This work is partially supported by Natural
Science Foundation of China(10575028) and  Natural Science
Foundation of Hebei Province(103143).

\newpage
\noindent\textbf{Figure caption}
\hspace{0.5cm}

Fig.1 The nuclear Drell-Yan cross section ratios
$R_{{A_1}/{A_2}}(x_2)$ on Fe to Be(left) and W to Be(right). Solid
curves correspond to nuclear effects on structure function . Dotted
and dash curves show the combination of HKM cubic type of nuclear
parton distributions with the linear energy loss $\alpha=1.27$ and
$1.99$, respectively. The experimental data are taken from the
E866[7].

\hspace{0.5cm}

Fig.2 The nuclear Drell-Yan cross section ratios
$R_{{A_1}/{A_2}}(x_2)$ on Fe to Be(left) and W to Be(right). Solid
curves correspond to nuclear effects on structure function . Dotted
and dash curves show the combination of HKM cubic type of nuclear
parton distributions with the quadratic energy loss $\beta=0.19$ and
$0.29$, respectively. The experimental data are taken from the
E866[7].

\hspace{0.5cm}

Fig.3 The nuclear Drell-Yan cross section ratios
$R_{{A_1}/{A_2}}(x_2)$ on W to D at 120GeV and 50GeV incident proton
beam with the linear energy loss $\alpha=1.27GeV/fm$(dash curves)
and $\alpha=1.99GeV/fm$(dotted  curves). Solid curves correspond to
nuclear effects on the structure function.

\hspace{0.5cm}

Fig.4 The nuclear Drell-Yan cross section ratios
$R_{{A_1}/{A_2}}(x_2)$ on W to D at 120GeV  incident proton beam
with the quadratic energy loss$\beta=0.19GeV/fm^2$(solid curve) and
the linear energy loss $\alpha=1.27GeV/fm$(dotted curve).

\newpage

\begin{figure}
\centering
\includegraphics[width=1.0\textwidth]{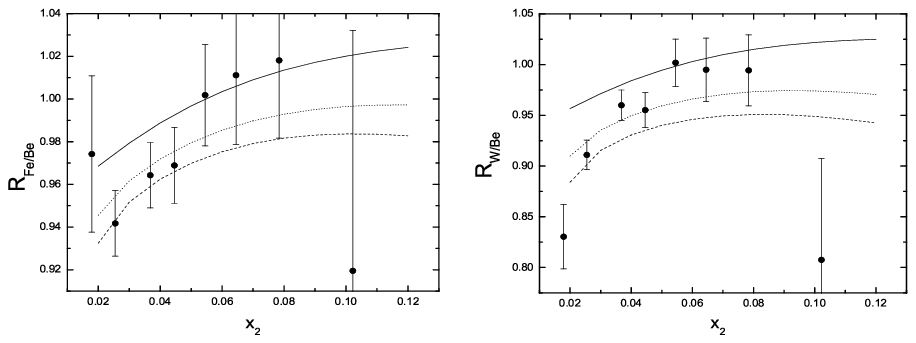}
\caption{}
\end{figure}

\begin{figure}
\centering
\includegraphics[width=1.0\textwidth]{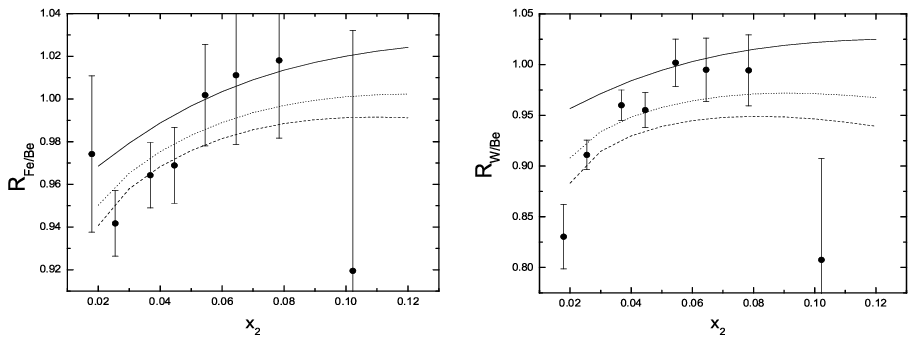}
\caption{}
\end{figure}

\begin{figure}
\centering
\includegraphics[width=1.0\textwidth]{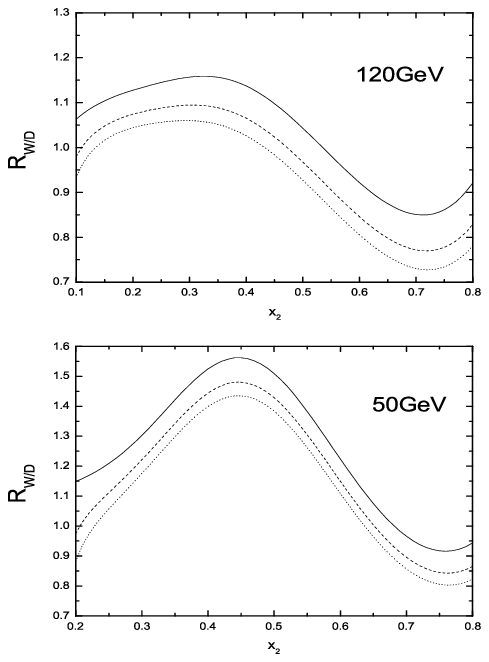}
\caption{}
\end{figure}

\begin{figure}
\centering
\includegraphics[width=1.0\textwidth]{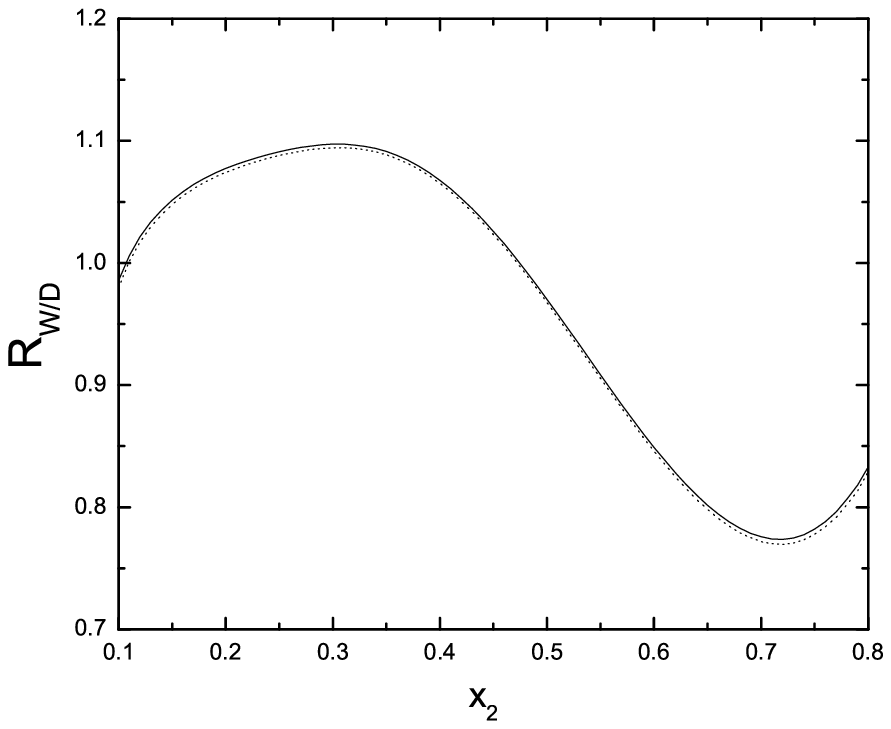}
\caption{}
\end{figure}


\end{document}